\documentclass[12pt,preprint]{aastex}

\begin{document}
\title{The black hole fundamental plane: revisit with a larger sample of
radio and X-ray emitting broad-line AGNs}
\author{Zhao-Yu Li, Xue-Bing Wu and Ran Wang}
\affil{Department of Astronomy, Peking University,
    Beijing 100871, China}
\email{zhaoyuli@pku.edu.cn; wuxb@bac.pku.edu.cn; littlestar@pku.edu.cn}

\begin{abstract}
We use a recently released SDSS catalog of X-ray emitting AGNs in
conjunction with the FIRST 20cm radio survey to investigate the
black hole fundamental plane relationship between the 1.4GHz radio
luminosity ($L_r$), 0.1-2.4 keV X-ray luminosity ($L_X$), and the
black hole mass ($M$), namely, $log L_r=\xi_{RX}log L_X+\xi_{RM}log
M +const$.  For this purpose, we have compiled a large sample of 725
broad-line AGNs, which consists of 498 radio-loud sources and 227
radio-quiet sources.  Our results are generally consistent with
those in our previous work based on a smaller sample of 115 SDSS
AGNs. We confirm that radio-loud objects have a steeper slope
($\xi_{RX}$) in the radio-X-ray relationship with respect to
radio-quiet objects, and the dependence of the black hole
fundamental plane on the black hole mass ($\xi_{RM}$) is weak. We
also find a tight correlation with a similar slope between the soft
X-ray luminosity and broad emission line luminosity for both
radio-loud and radio-quiet AGNs, which implies that their soft X-ray
emission is unbeamed and probably related to the accretion process.
With the current larger sample of AGNs, we are able to study the
redshift evolution of the black hole fundamental plane relation for
both radio-loud and radio-quiet subsamples. We find that there is no
clear evidence of evolution for radio-quiet AGNs, while for
radio-loud ones there is a weak trend where $\xi_{RM}$ decreases as
the redshift increases. This may be understood in part as due to the
observed evolution of the radio spectral index as a function of
redshift. Finally, we discuss the relativistic beaming effect and
some other uncertainties related to the black hole fundamental
plane. We conclude that, although introducing scatters to the
fundamental plane relation, Doppler boosting alone is not enough to
explain the observed steeper value of $\xi_{RX}$ in the radio-loud
subsample with respect to the radio-quiet ones. Therefore, the
significant difference of $\xi_{RX}$ between radio-loud and
radio-quiet sources is probably also due to the different physical
properties of the jets.
\end{abstract}


\section{INTRODUCTION}
Astrophysical black holes do not emit light directly, but they can
be probed by their gravitational influence on neighboring matter,
which produces observable signatures of black hole activity. The key
mechanism for a black hole to become active is the accretion process
(Frank, King \& Raine 2002), which is usually accompanied by a
relativistic jet.  Accretion disks and jets can produce photons from
radio to X-ray band. The radio emission is usually believed to
originate from the synchrotron radiation of the jet (Begelman,
Blandford \& Rees 1984), while the optical/UV emission mostly comes
from the multicolor black body radiation emitted from the accretion
disk (Shakura \& Sunyaev 1973), and the X-ray radiation is usually
associated with the inner-most region of the accretion disk, where
the temperature is the highest. In some cases, the contribution of
inverse Compton scattering from high energy electrons in a disk
corona is also needed to account for the observed power-law
spectrum in the X-ray band (Haardt \& Maraschi 1993). If jet
production is directly related to the accretion process, we would
expect a natural correlation between radio and X-ray luminosities
(Merloni, Heinz \& Di Matteo 2003; Heinz \& Sunyaev 2003; Falcke et
al. 2004).

The radio-to-X-ray correlation has long been studied in both
Galactic black hole (GBH) candidates and active galactic nuclei
(AGNs).  Gallo, Fender \& Pooley (2003) found a strong correlation
between the radio and X-ray emission ($L_r \propto L_X^{0.7}$) using
the simultaneous X-ray and radio observational data of stellar-mass
black hole X-ray binaries (XRBs) during the low/hard state. In
addition, they suggested that when XRBs enter the hard to soft
transition state, the jet is suppressed and the radio emission
decreases. Recently, some studies have shown that the substantial
scatter exists in such a relationship of GBHs (Gallo 2006; Xue \&
Cui 2007; Xue, Wu \& Cui 2008). In the case of AGNs, Brinkmann, Yuan
\& Siebert (1997) obtained a remarkable correlation between 2keV
X-ray luminosity and 5GHz radio luminosity for 324 radio-loud AGNs.
Canosa et al. (1999) found a strong correlation between soft X-ray
and 5GHz radio luminosities for 40 low-power radio galaxies.
Brinkmann et al. (2000) studied a sample derived from the
cross-correlation of the {\em ROSAT} All-Sky Survey (RASS) catalog
and the Very Large Array (VLA) Faint Images of the Radio Sky at
Twenty-Centimeters (FIRST 20cm) catalog. They found that for 843
AGNs, the X-ray and radio luminosities are correlated over two
decades in radio luminosity, spanning radio-loud and radio-quiet
regimes, but radio-quiet quasars seem to follow a different
correlation from radio-loud ones. Recently, Panessa et al. (2007)
investigated the radio/X-ray luminosity correlation for
low-luminosity AGNs, including local Seyfert galaxies and low
luminosity radio galaxies (LLRGs). They found that X-ray and radio
luminosities are significantly correlated over 8 orders of magnitude,
with Seyfert galaxies and LLRGs showing the similar slope, which seems
different from the previous results.

Theoretical explanations for the observed radio-to-X-ray relation
have been discussed in a number of previous works. Fender et al.
(2003) found that, at relatively lower accretion rate ( $\dot M <
7\times10^{-5}\dot M_{Edd}$), black hole X-ray binaries would enter
a ``jet-dominated'' state. At this stage, the majority of the
liberated accretion power is transferred into the jet and does not
dissipate as X-ray emission in the accretion flow. This was also
suggested by Falcke, K\"ording \& Markoff (2004), who demonstrated
that, below a critical value of the accretion rate ($<1\sim 10\%\dot
M_{Edd}$), the Spectral Energy Distribution (SED) of a black hole
accreting source is dominated by the non-thermal radiation from the
jet, while for sources with higher accretion rates ($\dot M \le \dot
M_{Edd}$), the radiation is dominated by the accretion flow. Using a
coupled disk-jet model, Yuan \& Cui (2005) showed that, in GBHs when
the X-ray luminosity is smaller than a certain value ($\sim <
10^{-5} -10^{-6} L_{Edd}$), the jet will dominate the radiation and
the radio-to-X-ray correlation becomes to $L_r \propto L_X^{1.23}$.
When the X-ray luminosity exceeds that critical value, the accretion
flow would produce most of the X-ray emission and the radio-to-X-ray
correlation then becomes to $L_r \propto L_X^{0.7}$. Therefore, the
accretion rate decides the detailed physical model of the accretion
disk, which then leads to different observed radio-to-X-ray
correlation slopes (K\"ording, Falcke \& Corbel 2006). Although it
is still not clear whether such a critical accretion rate is of the
same order of magnitude for both GBHs and AGNs, the similarity of
accretion-jet physics in these two systems seems to imply that
similar results may also be found in AGNs.

In a comprehensive study, Merloni et al. (2003) examined a sample
combining Galactic black hole systems and supermassive black hole
systems by investigating their compact emission in X-ray (2-10 keV)
and radio (5 GHz) bands. They found that the radio luminosity
($L_r$) is strongly correlated with both the black hole mass ($M$)
and the X-ray luminosity ($L_X$) (so called the fundamental plane of
black hole activity). The relation is: $log
L_r=\left(0.60_{-0.11}^{+0.11}\right)log
L_X+\left(0.78_{-0.09}^{+0.01}\right)log
M+\left(7.33_{-4.07}^{+4.05}\right)$. Subsequently, Wang et al.
(2006) selected a uniform sample of broad-line AGNs which was
cross-identified from the RASS, Sloan Digital Sky Survey (SDSS) and
FIRST 20cm radio survey to test the black hole fundamental plane
relation. Their final sample consisted of 115 broad emission line
AGNs including 39 radio-quiet AGNs and 76 radio-loud ones. They
found that the relationship has a very weak dependence on the black
hole mass. Moreover, radio-quiet and radio-loud objects have
different radio-to-X-ray slopes, which is 0.85 for radio-quiet
objects and 1.39 for radio-loud sources. This differs from the
result of Merloni et al. (2003) where the relationship seems to be
universal for different types of black hole sources. However, the
limited statistics of the sample in Wang et al. (2006) motivated us
to increase the number of sources in order to confirm the results on
stronger statistical bases. With a larger sample, we are also able
to study the possible evolution of the black hole fundamental plane
relation as we have enough sources in each redshift bins.

We organize the paper as the following. In $\S 2$ we present our
sample selection criteria and the main properties of our sample. In
$\S 3$ we show the derived fundamental plane relation and
investigate the possible evolution of such a relation. In $\S 4$ we
briefly discuss and summarize our results.

\section{THE RADIO AND X-RAY EMITTING BROAD-LINE AGN SAMPLE}
\subsection{Sample selection}

Our sample is selected based on the cross-identification of the
newly published X-ray emitting SDSS AGN catalog \citep{and07} and
the catalog of the FIRST 20cm radio survey\footnote{See VizieR
Online Data Catalog, 8071 (Becker et al. 2003)} \citep{whi97}.
\citet{and07} employed X-ray data from the {\em ROSAT} All-Sky Survey
(RASS) and both optical imaging and spectroscopic data from the
Sloan Digital Sky Survey (SDSS). It is worth emphasizing that the
RASS and SDSS are extremely well matched to each other via a variety
of coincidences (e.g., similar survey depth, sensitivity and sky
coverage, etc.)(Anderson et al. 2003). The RASS/SDSS data from $5740
\mathrm{\ deg^2}$ of sky spectroscopically covered in SDSS Data
Release 5 provide an expanded catalog of 7000 confirmed quasars and
other AGNs that are probable RASS identifications \citep{and07}.

One of the main benefits of this sample is that the SDSS surveyed
area is also covered by the FIRST 20cm radio survey. We
cross-correlate the broad-line AGN catalog of \citet{and07} with the
FIRST radio-detected sources to build a RASS-SDSS-FIRST
cross-identified sample of 868 broad-line AGNs. All of these 868
sources have been observed and detected at 1.4GHz (FIRST 20cm
survey) and in 0.1-2.4 keV energy band (RASS). The optical spectra
from the SDSS data archive can be used to estimate the central black
hole mass. Here, we exclude the high z sources (${z>2.171}$) in our
sample for the lack of the $\mathrm{H\beta}$ and \ion{Mg}{2}
$\lambda 2798$ emission lines in their SDSS optical spectra. Black
hole masses of the remaining 725 sources are estimated. The
radio-loudness ($R$) is calculated with the rest-frame B-band
(4400\AA) and 5GHz flux density according to the definition
$R=f_{5GHz}/f_B$ (Kellermann et al. 1989), and the 5GHz flux density
is derived from the 1.4GHz flux density assuming a spectral index of
0.5 (Kellermann et al. 1989). Radio-loud and radio-quiet sources are
separated by $R=10$. Please see subsection 2.2 of \citet{wang06}
 for the details of data reduction and black hole mass estimation
(Kaspi et al. 2000, 2005; McLure \& Javis 2002; Wu et al. 2004).

Our final sample comprises 725 entries, of which 498 are radio-loud,
227 are radio-quiet. In addition, we discover several radio-loud
narrow-line Seyfert 1 galaxies (NLS1) in our study. A detailed
discussion of these radio-loud NLS1s is presented in Appendix
A. Throughout this paper, we use the cosmology\footnote{We now
use this
   new cosmology to calculate the black hole mass from
  the H$\beta$ broad component's luminosity and FWHM.
  That is, $\mathrm{M_{bh}=1.464\times10^5\left(\frac{R_{BLR}}{1\ lt\ days}\right)
  \left(\frac{V_{FWHM}}{10^3\ km/s}\right)^2}$ and $\mathrm{Log R_{BLR}=(1.324\pm0.086)+(0.667\pm0.101)Log
  \left(\frac{L_{H\beta}}{10^{42}\ ergs\ s^{-1}}\right)}$. The latter relation
is an updated version of the $\mathrm{R_{BLR}-L_{H\beta}}$ relation
proposed by Wu et al. (2004) using the new cosmology.} with
$\mathrm{H_0=70\ km s^{-1} Mpc^{-1}}$,
 $\Omega_{\Lambda}=0.7$ and $\Omega_M=0.3$.


\subsection{Sample properties}

\begin{figure}
\epsscale{1.00} \plotone{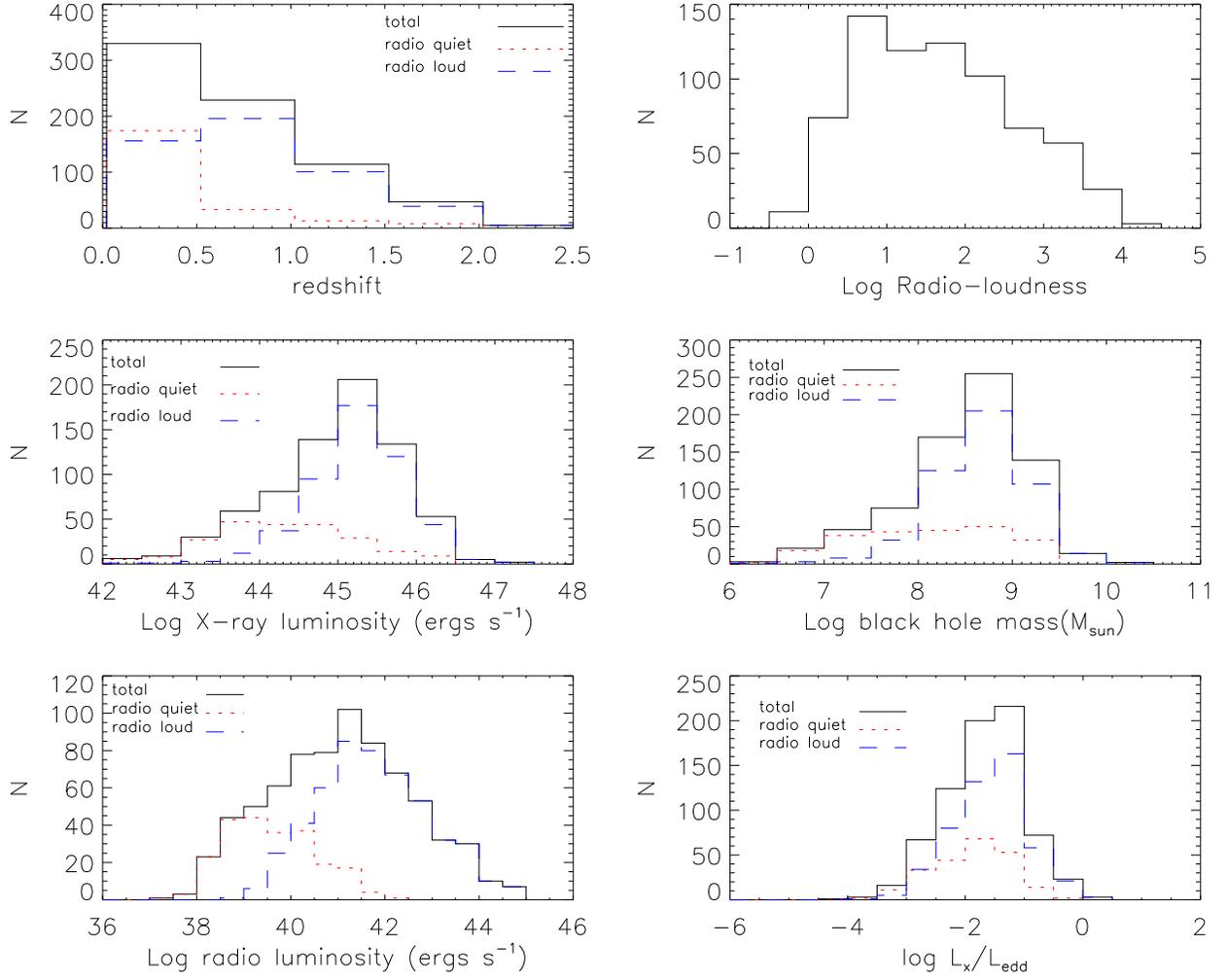} \caption{Global properties of our
AGN sample, with the top panels showing the histograms of the
redshift and logarithm of radio-loudness, the middle panels showing
the histograms of the X-ray luminosity and black hole mass (in units
of $M_\odot$), and the bottom panels showing the radio luminosity in
logarithm units and ratios of X-ray to Eddington
luminosity.\label{fig-1}}
\end{figure}

We describe here the main properties of our selected AGN sample.
Table \ref{tbl-1} gives the total 725 sources with the SDSS optical
name, redshift, broad-band (0.1-2.4 keV) soft X-ray luminosity,
rest-frame 1.4GHz radio luminosity, radio-loudness, black hole mass,
broad emission line luminosity and corresponding flag. In Figure
\ref{fig-1}, histograms of the redshift, logarithm of
radio-loudness, 0.1-2.4 keV X-ray luminosity, black hole mass (in
unit of $M_\odot$), 1.4GHz radio luminosity and ratio of X-ray to
Eddington luminosity ($L_{edd}$)\footnote{Defined as
$L_{Edd}=1.26\times 10^{38}(M/M_\odot)ergs\ s^{-1}$} of our sample
are plotted. The radio-loudness distribution does not show a clear
dichotomy between radio-loud and radio-quiet AGNs. This is not
surprised for samples of FIRST detected quasars. The
radio-loudness dichotomy was found mostly for optically selected
samples, and is usually absent for samples of deep radio-detected sources
(Brinkmann et al. 2000; White et al. 2000; Lacy et al. 2001).

Our sample is a part of the broad-line AGN catalog provided by
Anderson et al. (2007), which includes typical broad-line quasars,
Seyfert 1 galaxies, low redshift Seyfert 1.5-1.9 galaxies and some
rare galaxies like NLS1s. Compared with Wang et al.(2006), the
distribution ranges of different physical parameters in our sample
are relatively larger. The significant range in the luminosity at
each redshift bin can avoid the strong dependence of luminosity on
redshift in a flux-limited sample (Avni \& Tananbaum 1982). In
particular, more convincing results can be obtained with the
larger sample.

However, we must notice the selection biases in our sample. First of
all, AGNs with radio flux fainter than the FIRST detection limit can
not be detected by FIRST. Therefore, our sample does not include
radio-quiet AGNs with $\mathrm{log_{10}(Radio-Loudness)}
<-0.23$.\footnote{$10^{-0.23}$ is the minimum radio-loudness value
of radio-quiet AGNs in our sample} In addition, based on the
redshift distribution shown in Figure 1, it is clear that most AGNs
at high redshifts in our sample are radio-loud. Therefore, any conclusion
based on the total sample is biased toward luminous radio-loud AGNs.
We also perform the Kolmogorov-Smirnov (K-S) test to evaluate the
distribution similarity of physical parameters shown in Figure 1
between radio-quiet and radio-loud subsamples (such as
$\mathrm{L_r,\ L_X,\ M_{BH},\ and\ \frac{L_X}{L_{Edd}}}$). We found
that the probabilities of similarity are all less than 0.05, meaning
that the distribution of these physical quantities in the two
subsamples are essentially different, with relatively larger mean
value of each parameter for radio-loud AGNs.

\begin{figure}
\plotone{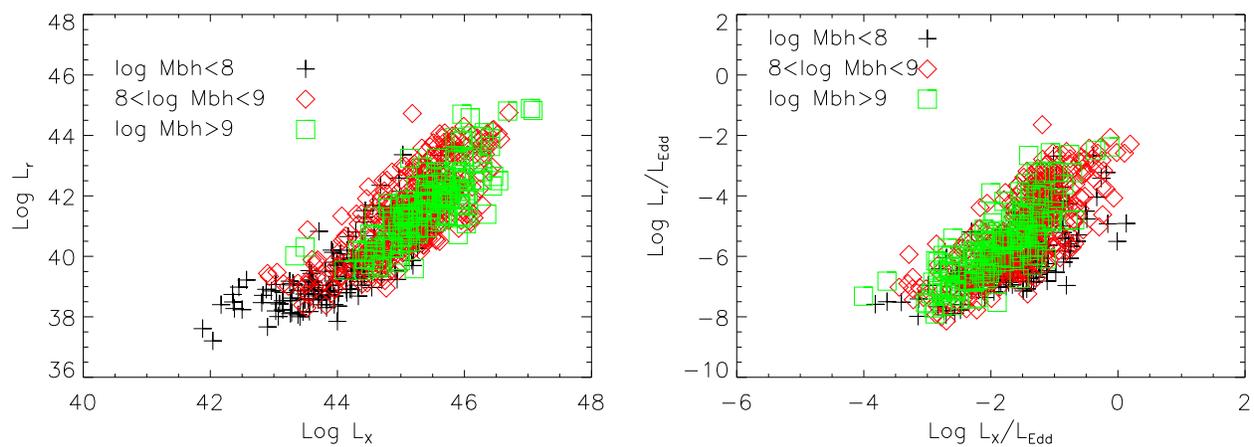} \caption{Rest-frame 1.4GHz radio luminosity
($L_r$) vs. the 0.1-2.4 keV X-ray luminosity, with different symbols
representing different logarithmic bins of the black hole mass (in
units of $M_\odot$). In the left panel we plot the logarithm of the
luminosity directly, while in the right panel we scale the radio and
X-ray luminosity with $\mathrm{\L_{Edd}}$. \label{fig-2}}
\end{figure}

We plot the radio vs. X-ray luminosity (left panel) and the
Eddington-luminosity-scaled radio vs. scaled X-ray luminosity (right
panel) in Figure \ref{fig-2}. Objects in different black hole mass
bins are presented with different symbols and colors to highlight a
possible segregation in the plot. Clearly, there is no trend that
sources in different black hole mass bins are parallel to each
other, which is consistent with the result in Wang et al. (2006).
Again, tight correlation between the radio and X-ray luminosities is
clear, with radio luminosity spanning more than 6 orders of
magnitude.

\begin{figure}
\plotone{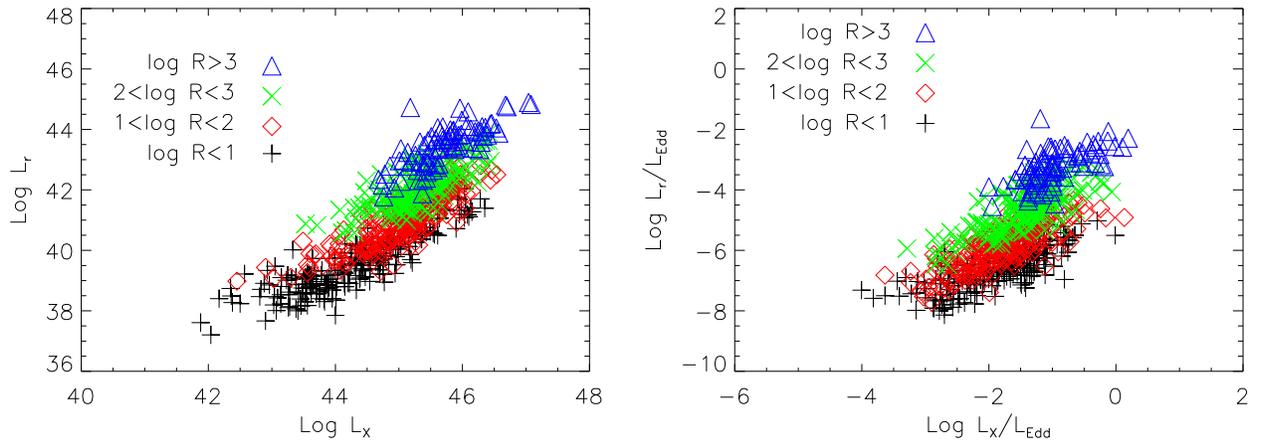} \caption{Rest-frame 1.4GHz radio luminosity
($L_r$) vs. the 0.1-2.4 keV X-ray luminosity, with different symbols
representing different radio-loudness bins. We plot
the radio vs. X-ray luminosity in the left panel, and the Eddington
luminosity scaled radio vs. X-ray luminosity in the right
panel.\label{fig-3}}
\end{figure}

Figure \ref{fig-3} and Figure 2 are identical except that in Figure
3, different colors and symbols are used to denote different
radio-loudness bins instead of black hole mass bins. AGNs in
different radio-loudness bins seem to distribute in parallel
sequences. The result also confirms the conclusion of
\citet{wang06}.


\section{CORRELATION ANALYSIS}
In this section, we derive the black hole fundamental plane relation
based on the current large sample. We also adopt several
statistical methods to test its significance. Finally, we examine
the possible evolution of this relation at different redshifts.

\subsection{The fundamental plane relation}
In the three-dimensional space ($\mathrm{logL_r, logL_X, logM}$),
black hole systems are preferentially distributed on a plane, called the
``fundamental plane of black hole activity'' \citep{mer03}. Based on
our sample of AGNs, we fit the data in the form:
\begin{equation}
\mathrm{ log\left(\frac{L_r}{10^{40}ergs\
s^{-1}}\right)=\xi_{RX}log\left(\frac{L_X}{10^{44}ergs\
s^{-1}}\right)+\xi_{RM}log\left(\frac{M}{10^8\
M_\odot}\right)+const.}
\end{equation}

We also fit the relation between Eddington-luminosity-scaled radio
and X-ray luminosities for the radio-quiet subsample only. The
fitting formula is
\begin{equation}
\mathrm{
log\left(\frac{L_r}{L_{Edd}}\right)=\xi_{ERX}log\left(\frac{L_X}{L_{Edd}}\right)+constant.}
\end{equation}

We apply the Ordinary Least Squares(OLS) multivariate regression
method (Isobe et al.1990) to the total, radio-loud and radio-quiet
subsamples, respectively. The OLS bisector fitting result for
equation (1) with errors at the one-sigma confidence level and the
dispersion ($\sigma_r$)\footnote{We define the dispersion as the
square root of the variance of the differences between the observed
radio luminosity and that calculated from the fitting relation} are
given in Table \ref{tbl-2}. We also list results of previous works
for comparison. Our result is generally consistent with Wang et al.
(2006), but now with smaller uncertainty for each coefficient due to
the larger sample. The conclusions are similar: first, the
coefficient $\xi_{RX}$ tends to be larger as the radio-loudness
increases; second, the black hole mass seems to be unimportant in
the correlation between radio and soft X-ray luminosities. We should
note that the black hole fundamental plane coefficients calculated
by Merloni et al. (2003) are based on a radio-loud and radio-quiet
combined sample, with a predominance of radio-quiet sources. Our
result for the radio-quiet subsample is similar to that of Merloni
et al. (2003)£¬ but the dependence on the black hole mass is weaker
in our case.

The fitting result for equation (2) with errors at the one-sigma
confidence level is
\begin{equation}
\mathrm{
log\left(\frac{L_r}{L_{Edd}}\right)=\left(0.96\pm0.04\right)log\left(\frac{L_X}{L_{Edd}}\right)+\left(-4.82\pm0.08\right).
}
\end{equation}
The dependence of $\mathrm{L_r/L_X}$ on the black hole mass is shown
in Figure 4. The overall correlation is weak, except that a positive
correlation is observed for sources (mostly radio-quiet ones) with
black hole masses smaller than $\mathrm{10^{7.5}M_\odot}$. Figure 5
shows the edge-on black hole fundamental plane relation for
radio-quiet sources, with radio-loud AGNs overplotted for
comparison. We also plot $\mathrm{L_r/L_{Edd}\ vs.\ L_X/L_{Edd}}$ in
Figure 6, with different symbols representing different
radio-loudness bins. Parallel sequences can be seen clearly from
these figures.

\subsection{Statistical tests for the fundamental plane relation}

\subsubsection{Partial correlation tests}
As \citet{bre05} pointed out, the correlation between X-ray and
radio luminosities may be dominated by the distance effect.
Following \citet{wang06}, we performed the partial Kendall $\tau$
correlation test to examine this effect \citep{akr96}. In Table
\ref{tbl-3}, Column (1)-(3) list the variable names of X, Y and Z
respectively, where the partial correlation of X and Y is calculated
with the influence of Z variable excluded. Column (4) gives the
subsample type. Column (5) lists the number of sources in each
subsample. Column (6)-(8) show results of the partial correlation
test, the square root of the calculation variance and the
probability of the null hypothesis. The null hypothesis is rejected
with a probability less than the significance
level(i.e.,$\sim0.05$). From Table \ref{tbl-3} we can see that the
partial correlation between $L_r$ and $L_X$ is strong even after
excluding the distance effect, because the $P_{null}$ value is less
than $10^{-10}$.

\subsubsection{The scrambling test}
Besides the partial correlation test performed above, we adopt
another method introduced by Bregman (2005) to evaluate the degree
of influence that any distance effect has on our sample (i.e.,
Merloni et al. 2006). We calculate the Pearson correlation
coefficient ($\rho$) between $Log\left(\frac{L_r}{10^{40} ergs\
s^{-1}}\right)$ and $\xi_{RX}Log\left(\frac{L_X}{10^{44} ergs\
s^{-1}}\right)+\xi_{RM}Log\left(\frac{M_{bh}}{10^8\ M_\odot}\right)$
by randomly assigning radio fluxes to objects in our sample. This
procedure is performed $10^6$ times in order to construct the Monte
Carlo test. The result is shown in Figure 7. The correlation
coefficients are adopted from Table 2. For comparison, we overplot
the Pearson correlation coefficient of the original sample with a
vertical line in each panel of Figure 7.

For radio-quiet objects, our $10^6$ realizations of randomized
datasets produce only one case, where the Pearson correlation
coefficient exceeds the value of the real datasets. For radio-loud
sources, \emph{not even one} shows a stronger correlation than the
real value. This means that, for the radio-quiet subsample, the
probability that the correlation of the fundamental plane is
entirely due to the distance effect is about $1\times10^{-6}$, and
it is even less than $10^{-6}$ for the radio-loud subsample and the
entire sample. Therefore, we are confident to say that the
existence of a correlation between the radio luminosity, X-ray
luminosity and black hole mass cannot be purely the result of
distance effects. This result is consistent with the partial
correlation test performed above.

\subsection{The correlation between soft X-ray and broad emission line luminosities}
In Figure 8, we plot the soft X-ray luminosity versus the H$\beta$
broad emission line luminosity (left panel) and the radio luminosity
versus the the H$\beta$ broad emission line luminosity (right panel)
in our sample. Radio-loud and radio-quiet sources are marked with
different symbols to highlight the possible dichotomy of the slope
in the figure. It is apparent in Figure 8 that radio-loud and
radio-quiet objects almost have the same slope in the left panel
($L_X$ vs. $L_{H\beta}$). The slope is $1.04\pm0.04$ for the
radio-quiet subsample and $1.00\pm0.04$ for the radio-loud
subsample. The Pearson correlation coefficients are 0.86 and 0.80,
respectively. However in the right panel of Figure 8 ($L_r$ vs.
$L_{H\beta}$), the slope of radio-loud sources is steeper than that
of radio-quiet ones. The slope is $1.68\pm0.08$ for radio-loud
sources and $0.99\pm0.03$ for radio-quiet sources, and the Pearson
correlation coefficients are 0.64 and 0.88, respectively.

The similar slope of radio-loud and radio-quiet sources in the $L_X$
vs. $L_{H\beta}$ plot seems to indicate that the soft X-ray emission
traces well the ionizing luminosity, and is probably isotropic and
closely related to the accretion process of the central black hole
for both radio-quiet and radio-loud sources. For radio-loud
broad-line AGNs, the jet contribution to the soft X-ray emission is
probably unimportant. The beaming effect due to the relativistic jet
seems also weak in the soft X-ray band. Otherwise, we should expect
the large scatter in the $L_X$ vs. $L_{H\beta}$ relation for
radio-loud sources in our sample.

The high redshift sources here usually have the measurement of the
\ion{Mg}{2} $\lambda$2798 broad emission line instead of the
H$\beta$ broad emission line in the observed wavelength. In Figure
9, we show the X-ray luminosity versus the \ion{Mg}{2} emission line
luminosity in our sample (left panel) and the radio luminosity
versus the \ion{Mg}{2} emission line luminosity (right panel).
The result is consistent with what derived from Figure 8 for low
redshift objects.

\subsection{Evolution of the black hole fundamental plane}

With the currently available large sample, we are able to
investigate the evolution of the black hole fundamental plane
relation by dividing the sample into different redshift bins. The
result is listed in Table 4 and plotted in Figure 10. For
radio-quiet objects, the coefficients $\xi_{RM}$ and $\xi_{RX}$ are
almost constant in different redshift bins. For radio-loud sources,
$\xi_{RX}$ is almost constant with redshift, while $\xi_{RM}$ seems
to decrease from positive to negative values as the redshift
increases, with the exception of the last redshift bin, where
$\xi_{RM}$ is positive again although it shows the largest error bars.

Now we can compare the evolution result obtained by us with some
theoretical predictions. From Heinz \& Sunyaev (2003), the radio
flux produced via the synchrotron radiation from relativistic jet
follows the scaling relation
\begin{eqnarray}
F_r \propto M^{\frac{2p+13-(2+p)\alpha_r}{8+2p}}{\dot
m^{\frac{2p+13+(p+6)\alpha_r}{2(p+4)}}}.
\end{eqnarray}
where $\alpha_r$ and $p$ represent the radio spectral index
($f_\nu\propto\nu^{-\alpha_r}$) and the electron energy distribution
index ($N\propto E^{-p}$) respectively. $M$ is the central black hole
mass. $\dot m$ is the dimensionless accretion rate ($\dot
m=\frac{\dot M}{\dot M_{Edd}}$). This relation is valid for
radio-loud AGNs, whose radio emission is believed to be produced via
the synchrotron radiation of the relativistic jet.

The dimensionless accretion rate ($\dot m$) is roughly proportional
to the ratio between broad emission line and Eddington luminosities
($\mathrm{\frac{L_{broad-line}}{L_{Edd}}}$)(Wandel, Peterson \&
Malkan 1999). In the subsection above, we have shown that the soft X-ray
luminosity is linearly scaled with the broad emission line
luminosity for both radio-loud and radio-quiet subsamples. Therefore,
we can use $L_X/L_{Edd}$ to represent the dimensionless accretion rate
($\dot m$). After replacing $\dot m$ with $L_X/L_{Edd}$ and
considering the relation between the Eddington luminosity and the
black hole mass ($L_{Edd}\propto M_{BH}$), equation (4) can be
turned into
\begin{eqnarray}
L_r \propto M^{-\alpha_r}{L_X^{\frac{2p+13+(p+6)\alpha_r}{2(p+4)}}}.
\end{eqnarray}
Therefore, the coefficient $\xi_{RM}$ and $\xi_{RX}$ of the radio-loud
subsample can be determined by $\alpha_r$ and $p$.

In our radio-loud AGN subsample, there are 114 sources that were
also detected by the Green Bank 4.85GHz northern sky survey (GB6)
(Gregore et al. 1996). Thus we are able to estimate the real radio
spectral index $\alpha_r$ ($f_{\nu} \propto {\nu}^{-\alpha_r}$) of
this small subsample with the observed flux densities at 1.4GHz and
4.8GHz. We plot the radio spectral index versus the redshift for
sources in this subsample in Figure 11. There is a weak trend that
the spectral index increases with the increasing of the redshift.
The mean values of $\alpha_r$ in each redshift bins are $-0.28$,
$-0.12$, $-0.31$, $-0.27$, $-0.06$, $-0.11$, and $0.04$ (from low-z
to high-z). We note that this is probably due to a selection effect
in observations, as we may miss high redshift AGNs with the negative
spectral index.

Using the average radio spectral indeces in different redshift bins,
we calculate the theoretical value of the coefficient $\xi_{RM}$ and
$\xi_{RX}$ based on equation (5). The general trend of
$\xi_{RM}$ is to decrease from 0.28 ($\alpha_r=-0.28$) to $-0.04$
($\alpha_r=0.04$), while $\xi_{RX}$ slightly increases from 1.23 to
1.44. This can be seen clearly from Figure~12. Therefore, the
evolution of $\xi_{RM}$ and $\xi_{RX}$ of the radio-loud subsample
with the redshift is consistent with the theoretical prediction.

On the other hand, it is also possible that the soft X-ray emission
of broad-line AGNs is produced by the synchrotron process in a hot
corona around accretion disks. Heinz (2004) calculated the
theoretical correlation between the radio luminosity, X-ray
luminosity and black hole mass when the jet radiation dominates. If
the physical parameters (i.e., magnetic field strength, electron
energy distribution index, etc.) of the hot corona and the
relativistic jet are similar, we can use the equations given by
Heinz (2004) to roughly estimate the coefficients $\xi_{RM}$ and
$\xi_{RX}$ in different redshift bins. We find that with this model,
$\xi_{RM}$ decreases from 0.047 ($\alpha_r=-0.28$) to -0.007
($\alpha_r=0.04$), and $\xi_{RX}$ increases from 1.23 to 1.44.
Therefore, even if the soft X-ray emission mechanism is synchrotron
radiation, the observed evolution of the black hole fundamental
plane coefficients can still be explained. We can not exclude such a
possibility because the origin of the soft X-ray emission is still
uncertain.

The radio spectral indeces of these 114 radio-loud AGNs obtained
from observations allow us to directly calculate their rest-frame
1.4GHz radio fluxes without assuming the canonical value of
$\alpha_r$ as 0.5. The black hole fundamental plane relation is
fitted based on this small subsample with measured $\alpha_r$. The
derived $\xi_{RX}$ and $\xi_{RM}$ coefficients are $1.34\pm0.16$ and
$-0.31\pm0.21$, respectively. These values are consistent within
errors with what we obtained from the radio-loud subsample if we set
$\alpha_r=0.5$ ($\xi_{RX}=1.50\pm0.08$ and $\xi_{RM}=-0.20\pm0.10$).
This indicates that our results for radio-loud AGNs are quite
robust.


\section{DISCUSSION AND CONCLUSION}

\subsection{Comparison with previous works}

The $\xi_{RM}$ coefficient we derived is rather small, meaning that
the dependence of the fundamental plane relation on the black hole
mass is weak, which is different from the result obtained by Merloni
et al. (2003). However, there are some differences between our
sample and that of Merloni et al. (2003). We only include the
broad-line AGN, while their sample includes both GBHs and SMBHs. In
our sample, the X-ray and radio emission are measured in 0.1-2.4keV
and 1.4GHz respectively, while Merloni et al. (2003) used the data
of 2-10keV X-ray emission and 5GHz radio core emission. It is still
unclear whether the soft (0.1-2.4keV) and hard X-ray (2-10keV)
emission have the similar origin for broad-line AGNs. A sample with
both available data in the soft and hard X-ray bands will help us
address this problem. This is beyond the scope of our current study
and will be done in the near future.

Another point worthy to mention is that the black hole mass data in
Merloni et al. (2003) were mainly obtained from the literatures.
This could introduce scatters due to the different mass measurement
techniques. In our current study, the black hole masses of AGNs were
estimated from the broad emission line and continuum properties by
the same method, which does not introduce any additional bias. In
addition, the black hole masses in our sample span a relatively
smaller range than that in Merloni et al. (2003) since we only
include broad-line AGNs.

\subsection{Relativistic beaming and other uncertainties}

Because the radio emission is produced by the relativistic jet,
Doppler boosting of the synchrotron radiation (namely the relativistic beaming)
would affect the observed radio flux significantly. Here we try to
address the question that whether the observed larger fundamental
plane coefficient $\xi_{RX}$ (or the larger radio luminosity) of radio-loud
sources is due to the Doppler boosting effect.

If the radio emission of both radio-loud and radio-quiet quasars are
from jets, the larger radio luminosity observed in radio-loud
sources can be considered to have much stronger Doppler boosting
effect than radio-quiet AGNs. In other words, we can assume that the
radio emission of radio-quiet AGNs is unbeamed. In subsection 3.3,
we already show that the soft X-ray emission of broad-line AGNs is
probably isotropic and unbeamed for both radio-loud and radio-quiet
sources. Therefore, for a radio-loud quasar, its {\em intrinsic}
radio luminosity ($L_{r,jet}$) (unbeamed) may be estimated with its
observed X-ray luminosity through the $L_r$-$L_X$ correlation
derived from radio-quiet sources. We will use the ratio between the
observed radio luminosity ($L_r$) and the {\em intrinsic} radio
luminosity ($L_{r,jet}$) to represent the boosting factor of
radio-loud sources. The equation of the Doppler boosting effect was
given by Heinz \& Merloni (2004):

\begin{eqnarray}
L_\nu=\frac{L_{\nu,jet}}{\Gamma^{k+\alpha_r}}\biggl[\frac{1}{(1+\beta
cos\theta)^{k+\alpha_r}}+\frac{1}{(1-\beta
cos\theta)^{k+\alpha_r}}\biggr].
\end{eqnarray}

In Figure 13, we show the distribution of the boosting factor
($L_r/L_{r,jet}$) of the radio-loud subsample (left panel) and the
boosting factor as a function of the inclination angle $\theta$
(right panel) when different Lorentz factor ($\Gamma$) is given. It
is apparent in Figure 13 that only with smaller $\theta$
($<5^\circ$) and larger $\Gamma$ ($>10$), we can produce boosting
factors as high as 1000. Such conditions can only be met in BL Lac
objects and are not likely to be the case for normal broad-line AGNs
studied in this work. With the typical Lorentz factor of
$\Gamma\sim5$ for broad-line AGNs (Orr \& Browne 1982) and a
non-negligible inclination angle ($\theta\gtrsim10^\circ$) (Maraschi
et al. 1994), the boosting value is estimated to be less than 30. As
can be seen in Figure 13, there are about half of radio-loud sources
whose boosting factors are larger than the predicted maximum
boosting value. Therefore, although the Doppler boosting effect
indeed has significant influence on the radio emission of radio-loud
AGNs, this effect alone is not enough to explain the observed larger
radio luminosity of radio-loud broad-line AGNs. This is consistent
with the result given by Heinz \& Merloni (2004) for the case of
unbeamed X-ray emission.

The non-simultaneous observations in the radio, X-ray and optical
bands for sources in our sample may lead to other uncertainties.
Both the ROSAT and FIRST surveys were conducted in the 1990s (Becker
et al. 1995; Voges et al. 1999; Britzen et al. 2007). SDSS-I
observations started from 1998, and ended in 2005 (York et al.
2000). So the data in different bands were obtained within ten years
or so. Unless the luminosities of most objects in our sample varied
significantly in these years, our result may not be affected too
much due to this effect. In addition, several (2$\sim$3) factors
change of the luminosities of some individual objects have little
influence on the statistically significant results derived here for
a large sample. However, as broad-line AGNs often show X-ray
variabilities in timescales from hours to decades, the
non-simultaneous observations could be an issue if the X-ray fluxes
of AGNs vary significantly. Therefore, simultaneous observations in
different bands for a sample of broad-line AGNs are still needed to
confirm our results, although they are difficult to conduct for a
large sample.

\subsection{Emission mechanisms}

The existence of the AGN radio-loud and radio-quiet dichotomy is
still an unsolved issue. There are evidences that the radio emission
of radio-quiet AGNs is likely produced by a weak (sub-relativistic)
jet near the black hole (Blundell \& Beasley 1998; Leipski et al.
2006), while radio-loud quasars are usually associated with large
scale jets of higher radio power (Rawlings \& Saunders 1991; Miller,
Rawlings \& Saunders 1993). Here we fit the black hole fundamental
plane relation for a broad-line AGN sample and find that the
coefficients are quite different between radio-loud and radio-quiet
quasars, especially the coefficient $\xi_{RX}$, which is steeper for
radio-loud objects. We also find that the larger radio luminosities
of radio-loud AGNs can not be produced by the Doppler boosting
effect alone. Therefore, the radio emission mechanism may be quite
different between radio-loud and radio-quiet AGNs. Such kind of
difference can be caused by many uncertain physical parameters, such
as the detailed disk/jet magnetic field strength and configuration,
the electron energy distribution, and the black hole spin,
etc.(Heinz \& Sunyaev 2003; Sikora et al. 2007). More detailed
observational and theoretical studies are still required.

For the soft X-ray emission, we have shown that the correlation
slopes between the soft X-ray luminosity and the broad emission line
luminosity are all around 1 for both radio-loud and radio-quiet
AGNs. Therefore, the soft X-ray emission is probably produced via
the accretion process near the central black hole, and it is
unbeamed and isotropic for both radio-loud and radio-quiet sources.
Otherwise larger scatters should exist in the $\mathrm{L_X\ vs.\
L_{H\beta}\ (L_{MgII})}$ relation. However, there are many uncertain
factors related to the origin of the soft X-ray emissions of AGNs
(i.e., the accretion flow, hot corona, and warm absorber, etc.). A
detailed study of the soft X-ray spectra for a larger sample of
broad-line AGNs may give us clues to understand the origin of the
soft X-ray emissions.

\subsection{Conclusion}

We revisited the fundamental plane relation of the black hole
activity based on a large broad-line AGN sample selected on the
basis of the cross-identification of the RASS, SDSS, and FIRST
catalogs. The results of our work confirm the main result of Wang et
al. (2006), namely, the black hole fundamental plane relation of the
radio-quiet subsample is different from that of the radio-loud
subsample; the coefficient $\xi_{RX}$ becomes larger as the
radio-loudness increases; the black hole mass seems unimportant in
the black hole fundamental plane relation. We also found that the
soft X-ray emission is most likely produced via the accretion
process of the central black hole for both radio-quiet and
radio-loud sources. In particular, for radio-loud sources, the jet
contribution to the soft X-ray emission seems unimportant and the
Doppler boosting of the relativistic jet is also weak in the soft
X-ray band. Moreover, by dividing the radio-loud and radio-quiet
samples into different redshift bins, we studied the evolution of
the fundamental plane relation. For radio-quiet sources, there seems
to be no clear evolution, while for radio-loud objects, the
correlation coefficient $\xi_{RM}$ tends to decrease as the redshift
increases. We found that the evolution of the radio spectral index
can help us at least partly understand such an evolution. Finally,
we briefly discussed the beaming effect and some other uncertainties
associated with the fundamental plane relation derived here. We
found that Doppler boosting effect indeed has significant influence
on the radio emission of radio-loud AGNs, but this effect alone is
not enough to explain the observed radio luminosity of the
radio-loud sources.

We thank Fukun Liu, Lei Qian, Xian Chen, Shuo Li, Da-Wei Xu and
Weimin Yuan for helpful discussions, and Eric Peng for checking the
English. We are also grateful to the anonymous referee for his/her
helpful comments. This work is supported by the NSFC grant
No.10525113, the RFDP grant 20050001026, the NCET grant
(NCET-04-0022) and the 973 Program No.2007CB815405. Funding for the
SDSS has been provided by the Alfred P. Sloan Foundation, the
Participating Institutions, the National Science Foundation, the
U.S. Department of Energy, the National Aeronautics and Space
Administration, the Japanese Monbukagakusho, the Max Planck Society,
and the Higher Education Funding Council for England. The SDSS Web
Site is http://www.sdss.org/. The SDSS is managed by the
Astrophysical Research Consortium for the Participating
Institutions. The Participating Institutions are the American Museum
of Natural History, Astrophysical Institute Potsdam, University of
Basel, University of Cambridge, Case Western Reserve University,
University of Chicago, Drexel University, Fermilab, the Institute
for Advanced Study, the Japan Participation Group, Johns Hopkins
University, the Joint Institute for Nuclear Astrophysics, the Kavli
Institute for Particle Astrophysics and Cosmology, the Korean
Scientist Group, the Chinese Academy of Sciences (LAMOST), Los
Alamos National Laboratory, the Max-Planck-Institute for Astronomy
(MPIA), the Max-Planck-Institute for Astrophysics (MPA), New Mexico
State University, Ohio State University, University of Pittsburgh,
University of Portsmouth, Princeton University, the United States
Naval Observatory, and the University of Washington.


\clearpage

\begin{appendix}
\section{RADIO-LOUD NARROW-LINE SEYFERT 1 GALAXIES IN OUR SAMPLE}

Narrow-Line Seyfert 1 (NLS1) galaxies are a sub-class of AGN
population. Their optical broad permitted emission lines are usually
narrower ($\mathrm{FWHM_{H\beta}<2000\ km/s}$ ) than that in normal
Broad-Line Seyfert 1 (BLS1) galaxies(Osterbrock \& Pogge 1985). The
NLS1s also show weak $\mathrm{[OIII]5007/H\beta_{total}}$ emission
and strong FeII emissions (Boroson \& Green 1992). Recently, Komossa
et al.(2006) argued that the classical criteria
($\mathrm{FWHM_{H\beta}<2000\ km/s}$) is not well defined or even
completely arbitrary. They suggested that $R_{4570}>0.5$ may be a
physically more meaningful criteria to distinguish the NLS1 from the
ordinary BLS1.\footnote{The optical Fe II strength, $R_{4570}$, is
the ratio of the Fe II complex between the rest wavelength $\lambda
4434$\AA \ and $\lambda 4684$\AA \ to the total $H\beta$ flux,
including the narrow component (Boroson \& Green 1992)}

The study of the physical mechanism of NLS1s is still ongoing. There
is growing evidence that most NLS1s are objects with low black hole
masses and high accretion rates, close to or even above the
Eddington accretion rate (Collin \& Kawaguchi 2004).

Radio-loud NLS1s are rare objects in NLS1 population (Komossa et al.
2006). In order to understand their radio properties, it is
important to expand the number of radio-loud NLS1s. Anderson et al.
(2007) roughly examined the optical broad permitted emission line of
AGNs in their catalog. They marked those objects that have
$\mathrm{FWHM_{H\beta}<2000\ km/s}$ as ``NLS1?'' in the comment
columns of their tables. Seventy-four of these objects have optical
spectra from SDSS and are detected in the FIRST radio survey, and
thus are included in our analysis in this paper. We calculated the
radio-loudness values for these objects, and found that five of them
are radio-loud. We list the properties of the five sources in Table
A1.

In a newly published work, Yuan et al. (2008) present a
comprehensive study of a sample of 23 radio-loud NLS1 galaxies.
Among those 23 sources, two radio-loud NLS1s are also discovered in
our present work, which are SDSS J144318.56+472556.7 and SDSS
J114654.28+323652.3. These two independent studies confirm the nature
of radio-loud NLS1s of these two objects.

\end{appendix}

\begin{deluxetable}{lccccccc}
\tabletypesize{\scriptsize} \tablecaption{The AGN
Samples\label{tbl-1}\tablenotemark{a}} \tablewidth{0pt}
\tablehead{\colhead{Name} & \colhead{z} &
\colhead{log($\mathrm{L_X/1\ erg\ s^{-1}}$)} &
\colhead{log($\mathrm{L_r/1\ erg\ s^{-1}}$)} & \colhead{log R} &
\colhead{log(M/$\mathrm{M_{\odot}}$)} &
\colhead{$\mathrm{L_{broad-line}}$}\tablenotemark{b} &
\colhead{flag}\tablenotemark{c}} \startdata
SDSS J$000608.04-010700.7$  & 0.949  &  45.60  &       41.13  &  1.345  &   8.71   &  43.47  &  0 \\
SDSS J$000710.01+005329.1$  & 0.316  &  44.95  &       39.79  &  0.549  &   9.07   &  42.90  &  1 \\
SDSS J$004319.73+005115.4$  & 0.308  &  44.64  &       39.61  &  0.506  &   9.42   &  42.90  &  1 \\
\tablenotetext{a}{Only 3 rows of the catalog are shown here. A
complete catalog will be provided in the elecronic version}
\tablenotetext{b}{The broad emission line luminosity. It is represented by
the $H_{\beta}$ broad component luminosity or the Mg II emission
line luminosity (when $H_{\beta}$ is unavailable).}
\tablenotetext{c}{1 means that $\mathrm{L_{H_{\beta}}}$ is used for
$\mathrm{L_{broad-line}}$, 0 means that $\mathrm{L_{Mg II}}$ is used
for $\mathrm{L_{broad-line}}$.}
\enddata

\end{deluxetable}

\begin{deluxetable}{lccrrr}
\tabletypesize{\scriptsize}\tablecaption{Derived Fundamental Plane
Relation\label{tbl-2}}
\tablewidth{0pt}\tablehead{\colhead{Subsample} & \colhead{Number} &
\colhead{$\xi_{RX}$} & \colhead{$\xi_{RM}$} & \colhead{Constant} &
\colhead{$\sigma_r$}}\startdata Total & 725 & $1.47\pm0.06$ &
$0.04\pm0.07$ & $-0.33\pm0.06$ & 0.83 \\
 Radio-quiet & 227 & $0.73\pm0.10$ &
$0.31\pm0.12$ & $-0.68\pm0.07$ & 0.42 \\
Radio-loud & 498 & $1.50\pm0.08$ & $-0.20\pm0.10$ & $0.05\pm0.10$ &
0.75 \\
\citet{mer03} & \nodata & $0.60\pm0.11$ & $0.78^{+0.11}_{-0.09}$ &
$7.33^{+4.05}_{-4.07}$ & 0.88 \\
Total\tablenotemark{a} & 115 & $1.33\pm0.15$ & $0.30\pm0.18$ &
$-0.40\pm0.14$ & 0.89 \\
Radio-quiet\tablenotemark{a} & 39 & $0.85\pm0.10$ & $0.12\pm0.13$ &
$-0.77\pm0.07$ & 0.38 \\
Radio-loud\tablenotemark{a} & 76 & $1.39\pm0.17$ & $0.17\pm0.21$ &
$-0.17\pm0.21$ & 0.77 \\
\enddata
\tablenotetext{a}{Results derived by \citet{wang06}}
\end{deluxetable}

\begin{deluxetable}{lccccccr}
\tabletypesize{\scriptsize} \tablecaption{Partial Correlation Test
For The Fundamental Plane Correlation\label{tbl-3}} \tablewidth{0pt}
\tablehead{\colhead{X} & \colhead{Y} &\colhead{Z} &\colhead{Type}
&\colhead{Number} &\colhead{$\tau$} &\colhead{$\sigma$}
&\colhead{$P_{null}$}} \startdata
        log Lx        &      log Lr        &      log D       &      Radio-loud   &      498   &     0.321   &    0.0264   &    $<$1.000E-10  \\
        log Lx        &      log Lr        &      log D       &      Radio-quiet  &      227   &     0.270   &    0.0424   &     1.916E-10  \\
        log Lx        &      log Lr        &      log D       &         Total     &      725   &     0.308   &    0.0220   &    $<$1.000E-10  \\
        log Lx        &      log Lr        &      log M       &      Radio-loud   &      498   &     0.567   &    0.0201   &    $<$1.000E-10  \\
        log Lx        &      log Lr        &      log M       &      Radio-quiet  &      227   &     0.578   &    0.0366   &    $<$1.000E-10  \\
        log Lx        &      log Lr        &      log M       &         Total     &      725   &     0.598   &    0.0188   &    $<$1.000E-10  \\
        log(Lx/Ledd)  &      log(Lr/Ledd)  &      log D       &      Radio-loud   &      498   &     0.472   &    0.0249   &    $<$1.000E-10  \\
        log(Lx/Ledd)  &      log(Lr/Ledd)  &      log D       &      Radio-quiet  &      227   &     0.517   &    0.0340   &    $<$1.000E-10  \\
        log(Lx/Ledd)  &      log(Lr/Ledd)  &      log D       &         Total     &      725   &     0.412   &    0.0197   &    $<$1.000E-10  \\
        log(Lx/Ledd)  &      log(Lr/Ledd)  &      log M       &      Radio-loud   &      498   &     0.572   &    0.0204   &    $<$1.000E-10  \\
        log(Lx/Ledd)  &      log(Lr/Ledd)  &      log M       &      Radio-quiet  &      227   &     0.553   &    0.0325   &    $<$1.000E-10  \\
        log(Lx/Ledd)  &      log(Lr/Ledd)  &      log M       &         Total     &      725   &     0.550   &    0.0160   &    $<$1.000E-10  \\

\enddata
\end{deluxetable}

\clearpage
\begin{deluxetable}{rrrr}
\tabletypesize{\scriptsize}\tablecaption{Derived Fundamental Plane
Relation In Different Redshift Bins\label{tbl-4}}
\tablewidth{0pt}\tablehead{\colhead{Redshift Range} &
\colhead{$\xi_{RX}$} & \colhead{$\xi_{RM}$} & \colhead{$N$}}
\startdata \cutinhead{Radio-Quiet subsample}
$z<0.13$ & $0.33\pm0.20$ & $0.23\pm0.19$ & 77 \\
$0.13<z<0.4$ & $0.35\pm0.23$ & $0.16\pm0.20$ & 81 \\
$0.4<z<1.0$ & $0.49\pm0.34$ & $0.12\pm0.34$ & 48 \\
$z>1.0$ & $0.55\pm0.69$ & $0.45\pm0.77$ & 21 \\
\cutinhead{Radio-Loud subsample}
$z<0.3$ & $0.63\pm0.24$ & $0.15\pm0.23$ & 58 \\
$0.3<z<0.45$ & $1.59\pm0.38$ & $-0.26\pm0.24$ & 67 \\
$0.45<z<0.65$ & $0.81\pm0.35$ & $-0.03\pm0.21$ & 86 \\
$0.65<z<0.80$ & $1.73\pm0.53$ & $-0.21\pm0.24$ & 71 \\
$0.80<z<1.0$ & $0.70\pm0.40$ & $-0.91\pm0.41$ & 64 \\
$1.0<z<1.3$ & $0.95\pm0.44$ & $-0.66\pm0.34$ & 77 \\
$z>1.3$ & $1.27\pm0.38$ & $0.11\pm0.37$ & 75 \\
\enddata
\end{deluxetable}

\begin{deluxetable}{lccccr}
\tabletypesize{\scriptsize}\tablecaption{Radio-Loud Narrow Line
Seyfert 1 Objects\label{tbl-5}}\tablenum{A1}
\tablewidth{0pt}\tablehead{\colhead{Name} & \colhead{z} &
\colhead{$\mathrm{FWHM_{H_{\beta}}/1\ km\ s^{-1}}$} &
\colhead{log($\mathrm{M_{BH}/M_{\odot}}$)} & \colhead{log R} &
\colhead{$R_{4570}$}} \startdata


SDSS J$144318.56+472556.7$\tablenotemark{a} & 0.703 & 1810.1 & 7.14
& 2.91 &
5.50 \\
SDSS J$114654.28+323652.3$ & 0.465 & 2374.3 & 7.43 & 1.98 &  1.59 \\

SDSS J$073320.84+390505.2$ & 0.664 & 2867.8 & 7.86 & 2.80 & 3.27 \\
SDSS J$154510.96+345246.9$ & 0.516 & 3269.7 & 8.13 & 1.34 & 0.74 \\
SDSS J$123304.05-003134.1$\tablenotemark{a} & 0.471 & 3297.7 & 7.81 & 2.32 & 2.18 \\
\enddata
\tablenotetext{a}{$H_{\beta}$ emission line is fitted with just one
Gaussian}

\end{deluxetable}

\clearpage


\clearpage
\clearpage
\clearpage

\begin{figure}
\plotone{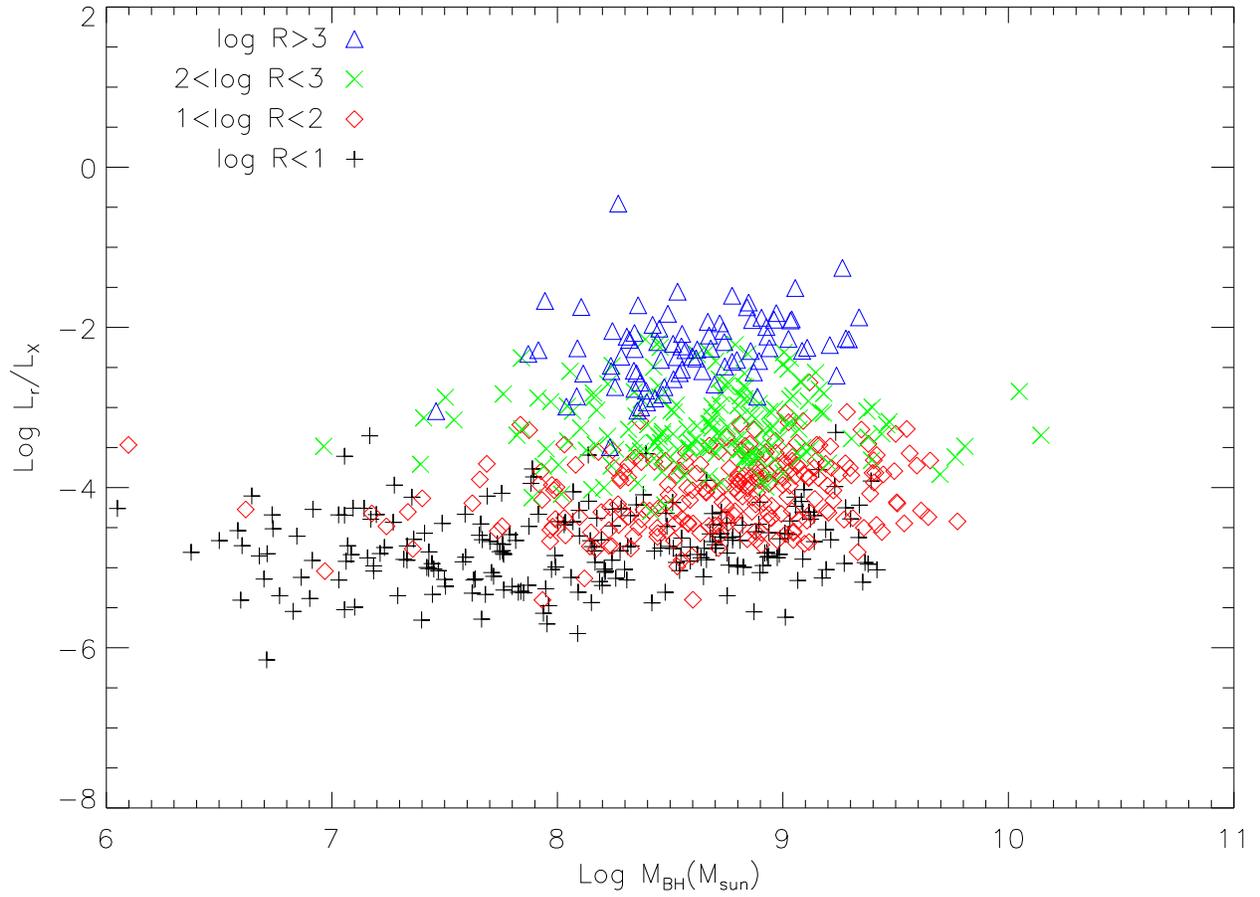} \caption{Ratio of the radio to X-ray luminosity
vs. the black hole mass. Different symbols represent objects in
different radio-loudness bins.\label{fig-4}}
\end{figure}
\clearpage

\begin{figure}
\plotone{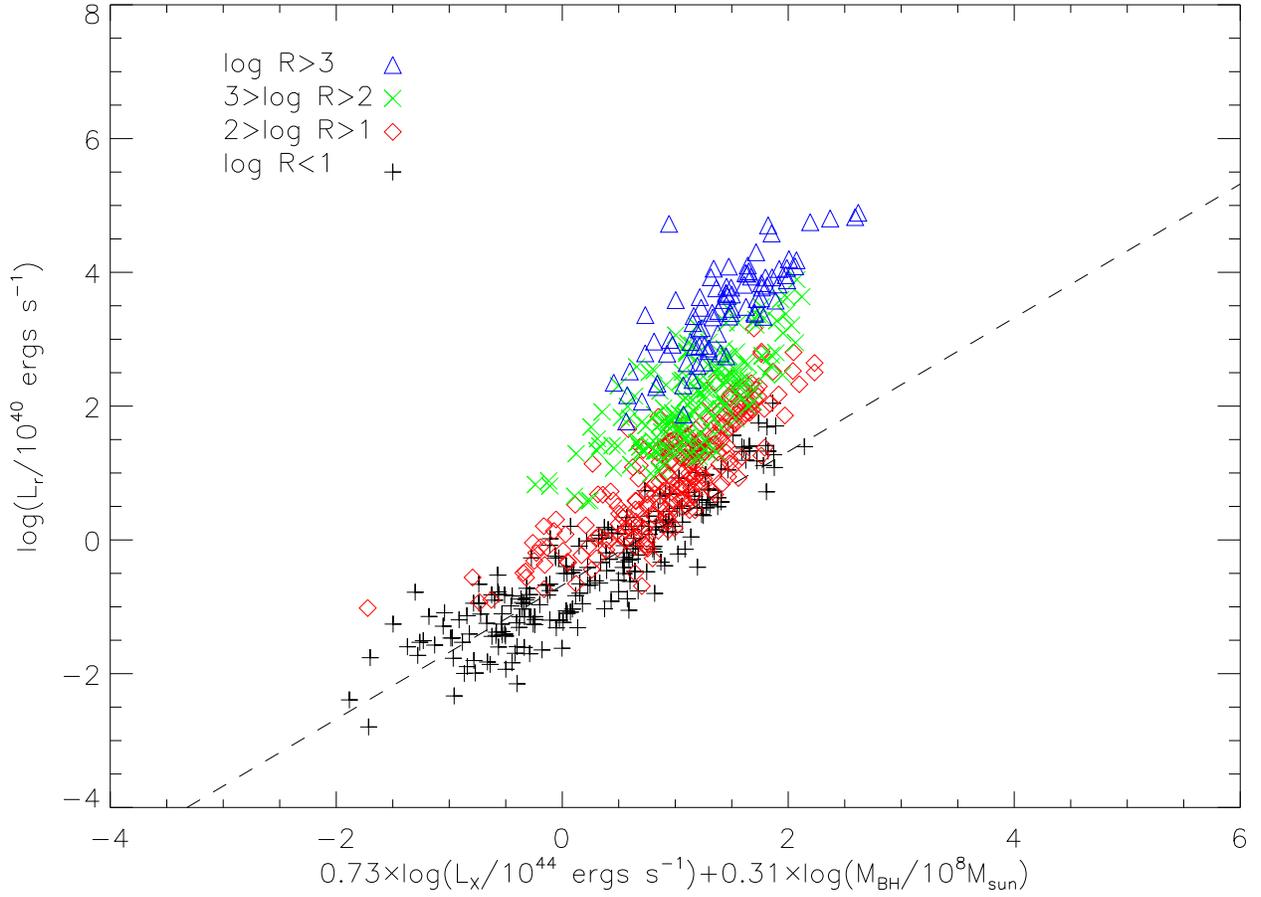} \caption{Edge-on view of the black hole
fundamental plane for all AGNs in our sample. Different symbols
represent sources in different radio-loudness bins. The dashed line
is the best-fitting result for radio-quiet AGNs.\label{fig-5}}
\end{figure}
\clearpage

\begin{figure}
\plotone{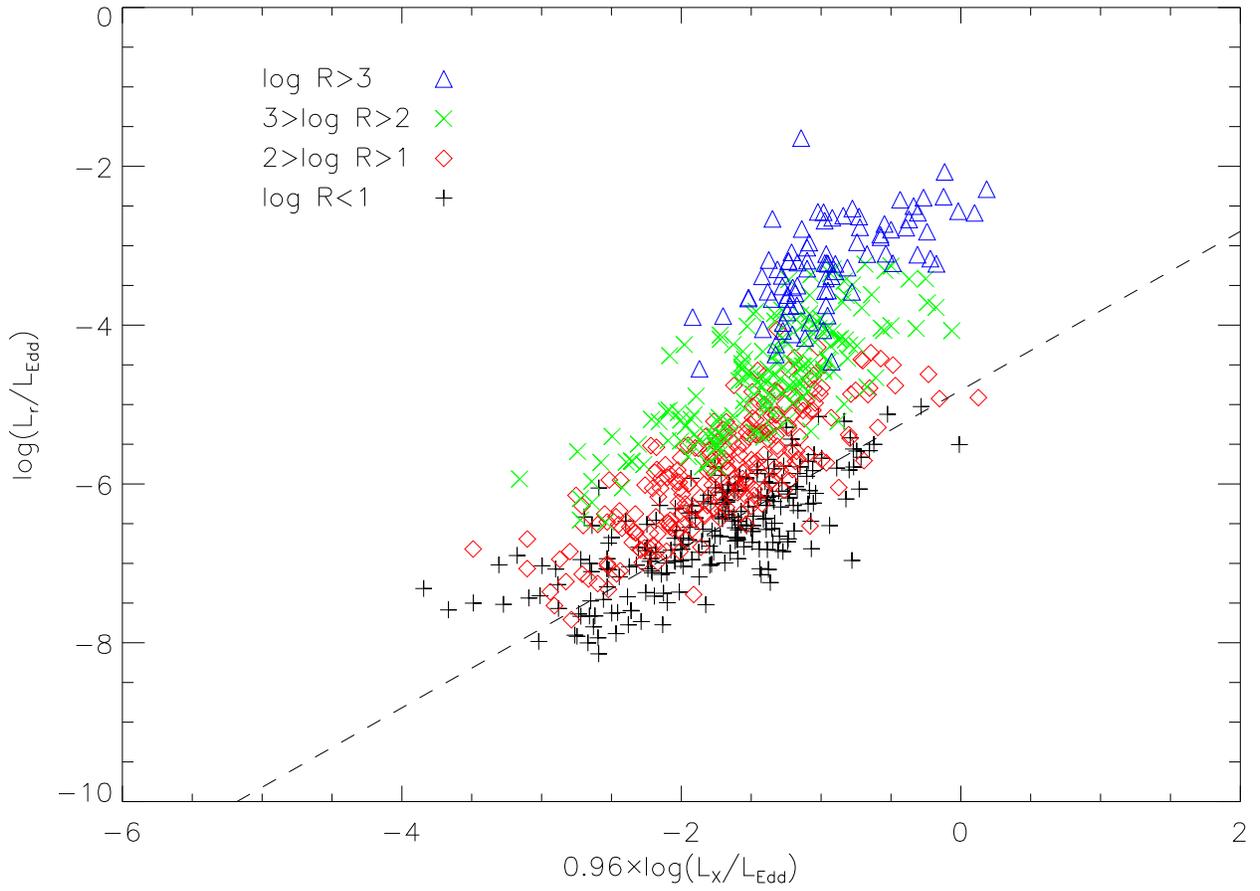} \caption{Correlation between $log(L_r/L_{Edd})$
and $0.96\times log(L_X/L_{Edd})$. The symbols and lines have
the same meanings as in Figure 5.\label{fig-6}}
\end{figure}
\clearpage

\begin{figure}
\epsscale{1.00} \plotone{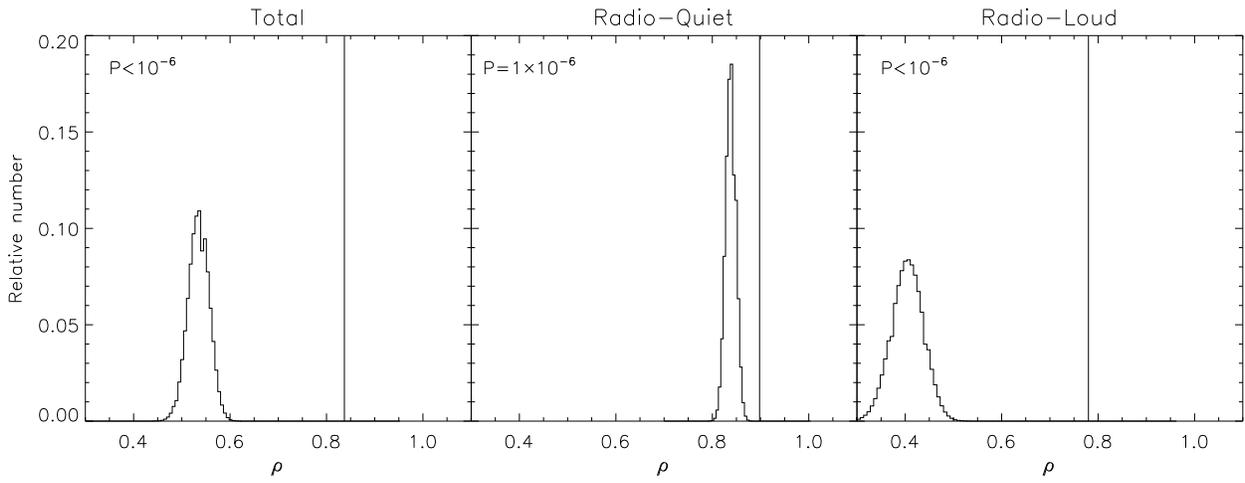} \caption{Results of the Monte
Carlo test using scrambled radio fluxes. The position of the
vertical line represents Pearson's correlation coefficient of the
real dataset.\label{fig-7}}
\end{figure}
\clearpage

\begin{figure}
\epsscale{1.00} \plotone{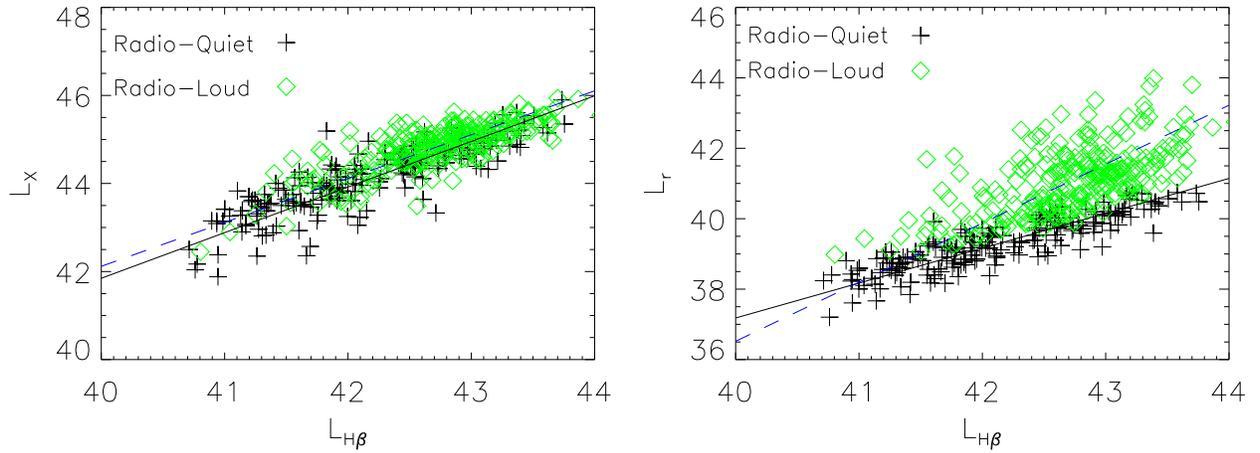} \caption{Plot of the 0.1-2.4keV
X-ray luminosity versus the H$\beta$ broad emission line luminosity
(left panel) and the rest-frame 1.4GHz radio luminosity versus the
H$\beta$ broad emission line luminosity (right panel). The meanings
of symbols and lines are: green diamond: radio-loud AGNs; black
cross: radio-quiet AGNs; black line: best-fitting result for the
radio-quiet subsample; dashed black line: best-fitting result for
the radio-loud subsample.\label{fig-8}}
\end{figure}
\clearpage
\begin{figure}
\epsscale{1.00} \plotone{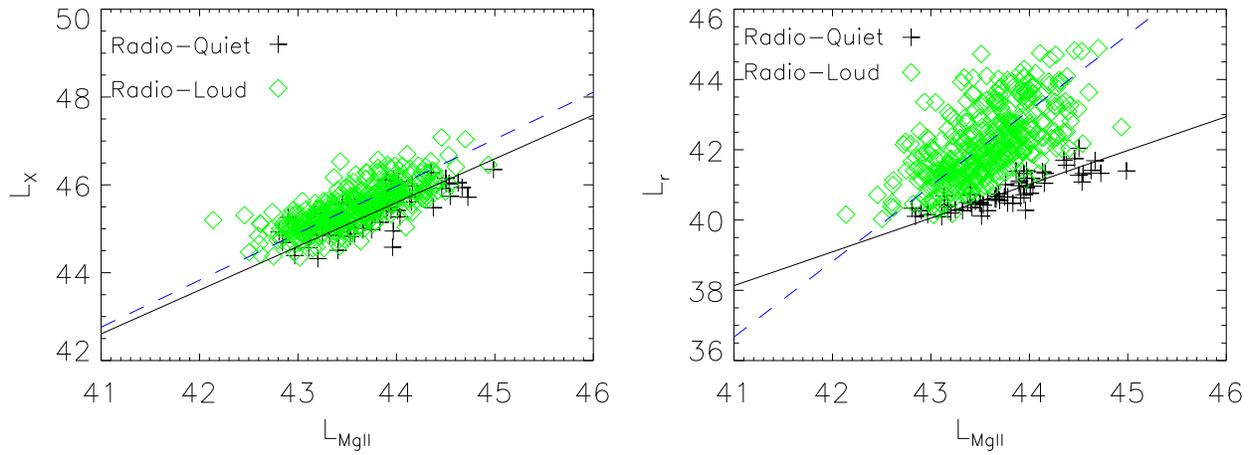} \caption{Plot of the 0.1-2.4keV
X-ray luminosity versus the \ion{Mg}{2} emission line luminosity
(left panel) and the rest-frame 1.4GHz radio luminosity versus the
\ion{Mg}{2} emission line luminosity (right panel). The symbols and
lines have the same meanings as in Figure 8.
\label{fig-9}}
\end{figure}
\clearpage

\begin{figure}
\epsscale{1.00} \plotone{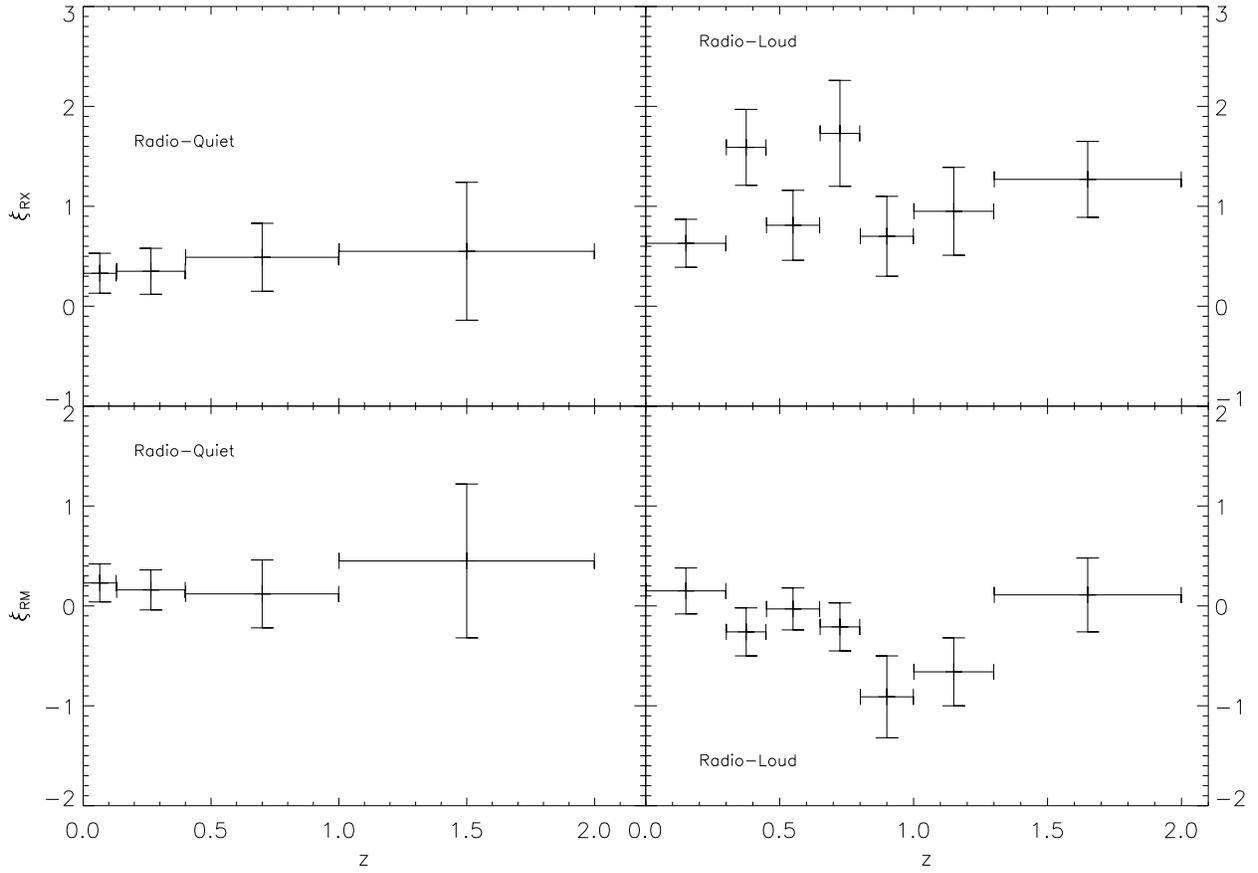} \caption{Dependence of the black
hole fundamental plane coefficients $\xi_{RM}$ and $\xi_{RX}$ on
redshift for radio-quiet and radio-loud AGNs.\label{fig-10}}
\end{figure}
\clearpage

\begin{figure}
\epsscale{1.00} \plotone{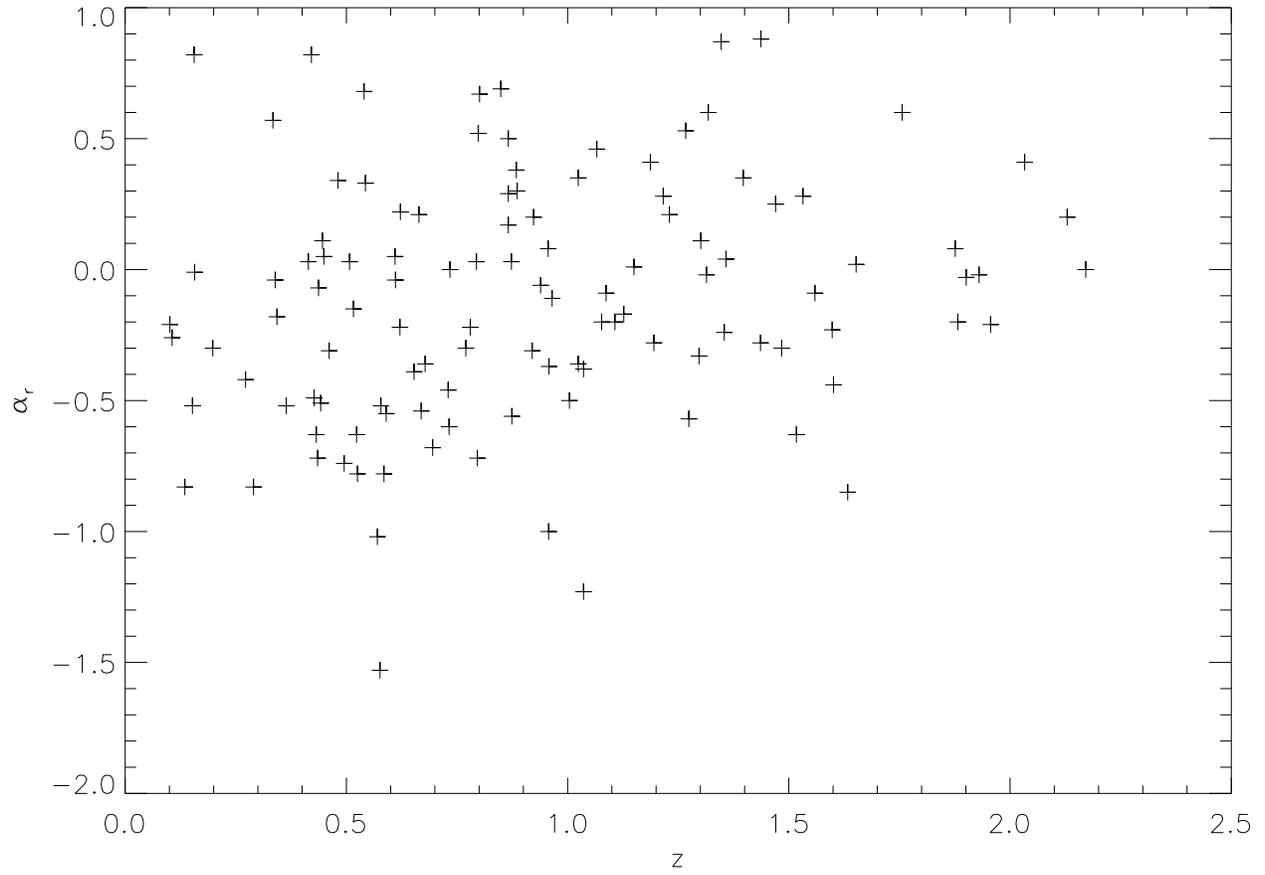} \caption{Redshift dependence of
the radio spectral index $\alpha_{r}$ of the 114 radio-loud AGNs
detected both at 1.4GHz and 5GHz frequencies.\label{fig-11}}
\end{figure}

\clearpage
\begin{figure}
\epsscale{1.00} \plotone{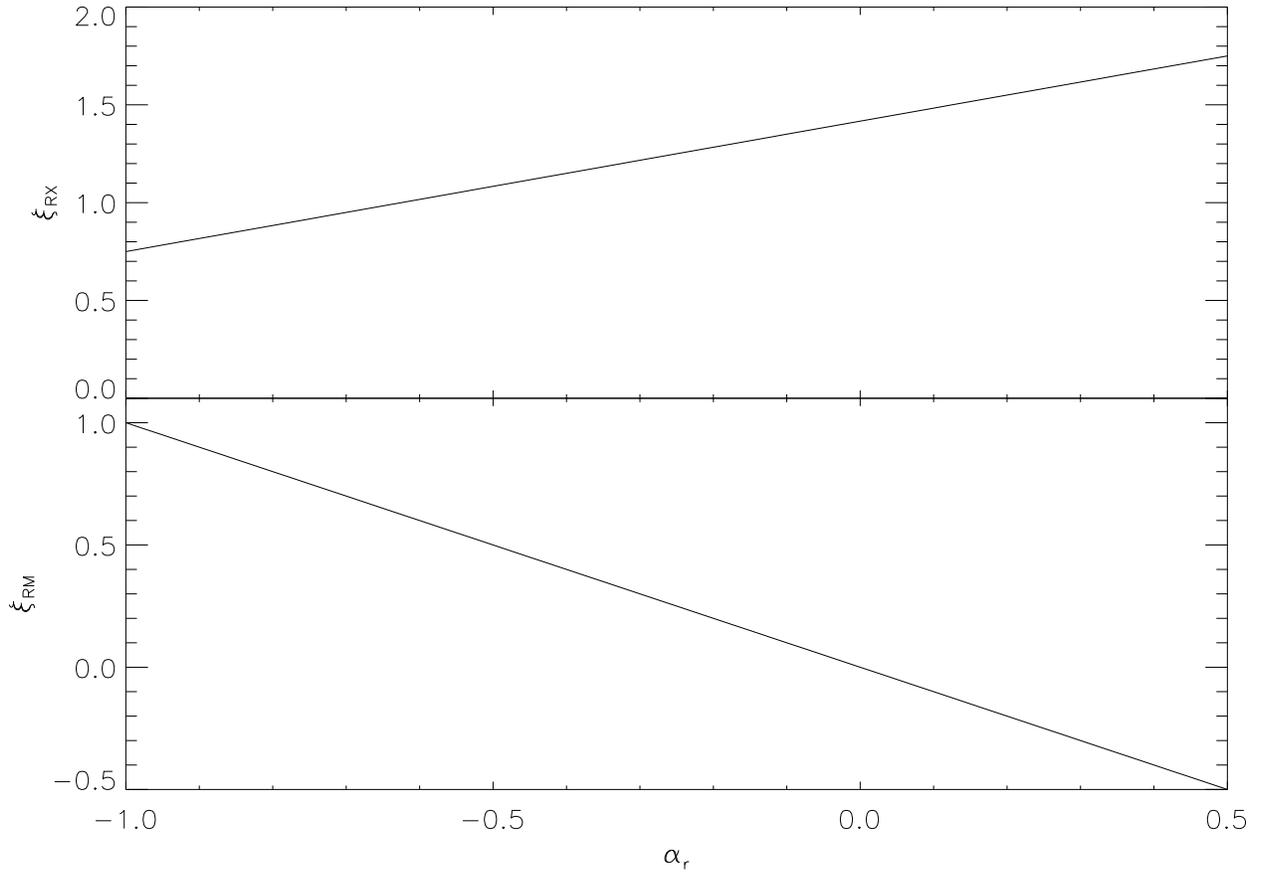} \caption{Predicted black hole
fundamental plane coefficients based on equation (5). Here,
$\alpha_{r}$ is the radio spectral index and $p$ equals to 2.
\label{fig-12}}
\end{figure}

\clearpage
\begin{figure}
\epsscale{1.00} \plotone{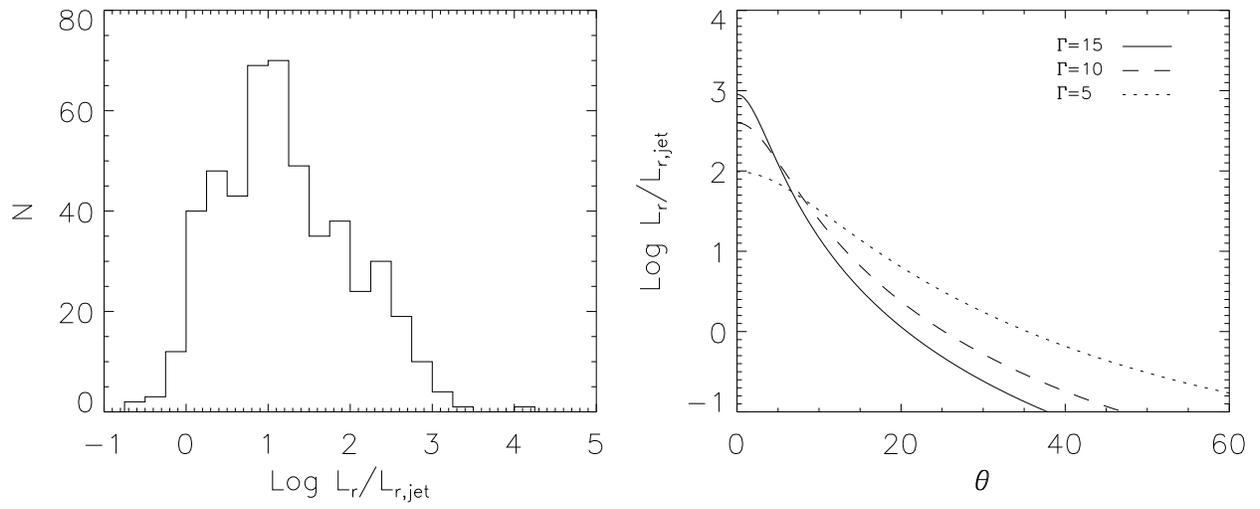}\caption{The histogram of the
boosting factor ($L_r/L_{r,jet}$) of the radio-loud subsample
(left panel) and the boosting factor as a function of the
inclination angle $\theta$ (right panel), with different line
representing different Lorentz factor ($\Gamma$).\label{fig-13}}
\end{figure}

\clearpage
\begin{thebibliography}{}
\bibitem[Akritas \& Siebert (1996)]{akr96} Akritas, M. G., \&
Siebert, J. 1996, \mnras, 278, 919
\bibitem[Anderson et al.(2003)]{and03} Anderson, S. F. et al.
2003, \aj,126,2009
\bibitem[Anderson et al.(2007)]{and07} Anderson, S. F. et al.
2007, \aj,133,313
\bibitem[Avni \& Tananbaum (1986)]{avn06} Avni, Y., \& Tananbaum, H.
1982, \apj, 262, L17
\bibitem[Becker et al.(2003)]{bec03} Becker, R. H., Helfand, D. J.,
White, R. L., Gregg, M. D., \& Laurent-Muehleisen, S. A. 2003,
VizieR Online Data Catalog, 8071
\bibitem[Becker et al. (1995)]{bec95} Becker, R. H., White, R. L.,
\& Helfand, D. J. 1995, \apj, 450, 559
\bibitem[Begelman, Blandford \& Rees (1984)]{beg84} Begelman, M. C.,
Blandford, R. D., \& Rees, M. J., 1984, Rev. Mod. Phys. 56, 255
\bibitem[Blundell \& Beasley (1998)]{blu98} Blundell, K. M., \&
Beasley, A. J. 1998, \mnras, 299, 165
\bibitem[Boroson \& Green (1992)]{bg92} Boroson, T. A., \& Green, R. F. 1992, \apjs, 80, 109
\bibitem[Bregman et al.(2005)]{bre05} Bregman, J. N. 2005, \apjl,
astro-ph/0511368
\bibitem[Brinkmann et al. (2000)]{bri00} Brinkmann W., Laurent-Muehleisen S. A., Voges W.,
Siebert J., Becker R. H., Brotherton M. S., White R. L., Gregg M. D.
2000, \aap, 356, 445
\bibitem[Brinkmann et al. (1997)]{bri97} Brinkmann W., Yuan W., \&
Siebert J., 1997, A\&A 319, 413
\bibitem[Britzen et al. (2007)]{bri07} Britzen, S. et al. 2007,
\aap, 476, 759
\bibitem[Canosa et al. (1999)]{can99}Canosa, C.M., Worrall, D.M.,
Hardcastle, M.J., \& Birkinshaw, M. 1999, \mnras, 310, 30
\bibitem[Collin \& Kawaguchi(2004)]{col04} Colin, S., \& Kawaguchi,
T. 2004, \aap, 426, 797
\bibitem[Falcke et al. (2004)]{fal04} Falcke, H., K\"ording, E., \& Markoff, S. 2004, \aap,
414, 895
\bibitem[Fender et al. (2003)]{fen03} Fender, R. P., Gallo, E., \& Jonker, P. G. 2003,
\mnras, 343, L99
\bibitem[Frank, King \& Raine (2002)]{fra02} Frank, J., King, A., \&
Raine, D., 2002, Accretion Power in Astrophysics, 3rd Edition,
Cambridge University Press
\bibitem[Gallo et al. (2003)]{gal03} Gallo,E., Fender, R. P., \& Pooley, G. G. 2003,
\mnras, 344, 60
\bibitem[Gallo et al. (2006)]{gal06} Gallo, E.,2006, AIP Conference
Proceedings, Vol. 924, 715
\bibitem[Gregory et al. (1996)]{gre96} Gregory, P. C., Scott, W. K.,
Douglas, K., \& Condon, J. J. 1996, \apjs, 103,427
\bibitem[Haardt \& Maraschi (1993)]{hm93} Haardt, F. \& Maraschi, L. 1993,
\apj, 413, 507
\bibitem[Heinz \& Merloni (2004)]{hei04} Heinz, S., \& Merloni, A.
2004, \mnras, 355, L1
\bibitem[Heinz et al. (2003)]{hei03} Heinz, S., \& Sunyaev, R. A. 2003, \mnras, 343, L59
\bibitem[Heinz (2004)]{hei04} Heinz, S. 2004, \mnras, 355, 835
\bibitem[Isobe et al.(1990)]{iso90} Isobe, T., Feigelson, E.D., Akritas, M.G.
\& Babu, G.J. 1990, \apj, 364, 104
\bibitem[Kaspi et al.(2000)]{kas00} Kaspi, S., Smith, P. S., Netzer,
H., Peterson, B. M., Vestergaard, M., \& Giveon, U. 2000, \apj, 533,
631
\bibitem[Kaspi et al.(2005)]{kas05} Kaspi, S., Maoz, D., Netzer, H.,
Peterson, B. M., Vestergaard, M., \& Jannuzi, B. T. 2005 \apj, 629,
61

\bibitem[Kellermann et al.(1989)]{kel89} Kellermann, K. I., Sramek,
R., Schmidt, M., Shaffer, D. B., \& Green, R. 1989, \aj, 98, 1195
\bibitem[Komossa et al.(2006)]{kom06} Komossa, S., Voges, W., Xu,
D., Mathur, S., Adorf, H. M., Lemson, G., Duschl, W. J., \& Grupe,
D. 2006, AJ, 132, 531
\bibitem[K\"ording, Falcke \& Corbel (2006)]{kor06}K\"ording, E., Falcke
H. \& Corbel S. 2006, \aap, 456, 439
\bibitem[Lacy et al. (2001)]{lac01} Lacy, M., Laurent-Muehleisen,
S. A., Ridgway, S. E., Becker, R. H., \& White, R. L. 2001, \apj,
551L, 17L
\bibitem[Leipski et al. (2006)]{lei06} Leipski, C., Falcke, H.,
Bennert, N., \& Huttemeister, S. 2006, \aap, 455, 161
\bibitem[Maraschi et al. (1994)]{mar94} Maraschi, L., \& Rovetti, F.
1994, \apj, 436, 79
\bibitem[McLure \& Jarvis(2002)]{mcl02} McLure, R. J., \& Jarvis,
M. J. 2002, \mnras, 337, 109
\bibitem[Merloni et al.(2003)]{mer03} Merloni, A., Heinz, S., \& Di
Matteo, T. 2003 \mnras, 345, 165
\bibitem[Merloni et al.(2006)]{mer06} Merloni, A., Kording, E.,
Heinz, S., Markoff, S., DiMatteo, T., \& Falcke, H. 2006, New
Astronomy 11, 567
\bibitem[Miller, Rawlings, \& Saunders (1993)]{mil93} Miller, P.,
Rawlings, S., \& Saunders, R. 1993, \mnras, 263, 425
\bibitem[Orr \& Browne (1982)]{orr82} Orr, M. J. L., \& Browne, I.
W. A. 1982, \mnras, 200, 1067
\bibitem[Osterbrock \& Pogge (1985)]{ost85} Osterbrock, D. E., \& Pogge,
R.W. 1985, \apj, 297, 166
\bibitem[Panessa et al. (2007)]{pan07} Panessa, F., Barcons, X., Bassani, L.,
Cappi, M., Carrera, F. J., Ho, L. C., \& Pellegrini, S. 2007, A\&A,
467, 519
\bibitem[Rawlings \& Sanuders (1991)]{raw91} Rawlings, S., \&
Saunders, R. 1991, Nature, 349, 138
\bibitem[Shakura \& Sunyaev (1973)]{sha73} Shakura, N. I.  \& Sunyaev, R.A.
1973, \aap, 24, 337
\bibitem[Sikora et al. (2007)]{sik07} Sikora, M., Stawarz, L., \&
Lasota, J. P. 2007, \apj, 658, 815
\bibitem[Voges et al. (1999)]{vog99} Voges, W. et al. 1999, \aap,
349, 389
\bibitem[Wandel, Peterson, \& Malkan (1999)]{wan99} Wandel, A.,
Peterson, B. M., \& Malkan, M. A. 1999, \apj, 526, 579
\bibitem[Wang et al.(2006)]{wang06} Wang, R., Wu, X-B., Kong, M. Z.
2006, \apj, 645, 890
\bibitem[White et al.(1997)]{whi97} White, R. L., Becker, R. H.,
Helfand, D. J.,\& Gregg, M. D. 1997, \apj, 475, 479
\bibitem[White et al. (2000)]{whi00} White, R. L. et al. 2000, \apjs, 126,
133
\bibitem[Wu et al.(2004)]{wu04} Wu, X-B., Wang, R., Kong, M. Z.,
Liu, F. K., \& Han, J. L. 2004, \aap, 424, 793
\bibitem[Xue \& Cui (2007)]{xue07} Xue, Y. Q., \& Cui, W. 2007, A\&A,
466, 1053
\bibitem[Xue, Wu \& Cui (2008)]{xue08} Xue, Y. Q., Wu, X.-B., \& Cui, W. 2008, \mnras, 384, 440
\bibitem[York et al. (2000)]{yor00} York, D. G. et al. 2000, \aj,
120, 1579
\bibitem[Yuan et al. (2005)]{yua05} Yuan, F., \& Cui, W. 2005, \apj, 629, 408
\bibitem[Yuan et al. (2008)]{yuan08} Yuan, W., Zhou, H. Y.,
Komossa, S., Dong, X. B., Wang, T. G., Lu, H. L., \& Bai, J. M.
2008, \apj, in press (astro-ph/0806.3755)
\end{thebibliography}
\end{document}